\documentclass[final,3p,times]{elsarticle}
\usepackage{lipsum}
\makeatletter
\def\ps@pprintTitle{%
 \let\@oddhead\@empty
 \let\@evenhead\@empty
 \def\@oddfoot{}%
 \let\@evenfoot\@oddfoot}
 
\usepackage{amssymb}
\usepackage{amsmath}
\usepackage{xurl}
\usepackage{hyperref}
\usepackage{subcaption}

\usepackage{orcidlink}
\usepackage{lmodern}
\usepackage{mathpazo}
\usepackage{graphicx}
\usepackage{algorithm}
\usepackage{algpseudocode}
\usepackage{booktabs}

\journal{Computer Methods in Applied Mechanics and Engineering}

\begin{document}

\begin{frontmatter}

\title{3D neuron growth and neurodevelopmental disorder modeling based on truncated hierarchical B-splines with multi-level local refinements}

% Use letters for affiliations, numbers to show equal authorship (if applicable) and to indicate the corresponding author
\author[a]{Kuanren Qian}
\author[a,b,c]{Yongjie Jessica Zhang}

\address[a]{Department of Mechanical Engineering, Carnegie Mellon University, 5000 Forbes Ave, Pittsburgh, PA 15213, USA}
\address[b]{Department of Biomedical Engineering, Carnegie Mellon University, 5000 Forbes Ave, Pittsburgh, PA 15213, USA}
\address[c]{Department of Civil and Environmental Engineering, Carnegie Mellon University, 5000 Forbes Ave, Pittsburgh, PA 15213, USA}

\begin{abstract}
3D neuron growth and neurodevelopmental disorders (NDDs) deterioration exhibit complex morphological transformations as neurites differentiate into axons and dendrites, forming intricate networks driven by tubulin concentrations and neurotrophin signals. 
Conventional 2D models fall short of capturing such morphological complexity, prompting the need and development of advanced 3D computational approaches. 
In this paper, we present a complex 3D neuron growth model based on isogeometric analysis (IGA) and the phase field method, utilizing locally refined truncated hierarchical B-splines (THB-splines). 
IGA offers isoparametric representation and higher-order continuity, which are essential for simulating the smooth, evolving interfaces of phase field neurites. 
In contrast, the phase field method can automatically handle diffuse interfaces and complex topological changes without explicit boundary tracking. 
This IGA-based phase field method enables accurate and efficient simulation of neurite extensions, branching, and retraction in a fully 3D setting. 
The THB-spline implementation supports multi-level local refinement, focusing computational resources on regions of active growth, while dynamic domain expansion adapts the simulation domain to extend with growing neurites. 
KD-tree-based interpolation ensures that phase field variables are accurately transferred onto newly refined meshes. 
NDDs associated neurite deterioration is simulated by modulating the driving force term within the phase field model to induce interface retraction.
This comprehensive 3D framework enhances the accuracy of neurite morphology simulations, advancing the study of complex neuron development, network formation and NDDs.
\end{abstract}

\begin{keyword}
Neuron growth \sep Neurodevelopmental disorder \sep Isogeometric analysis \sep Truncated hierarchical B-splines \sep Local refinement \sep Adaptive domain expansion \sep Phase field method
 %% PACS codes here, in the form: \PACS code \sep code
%% MSC codes here, in the form: \MSC code \sep code
%% or \MSC[2008] code \sep code (2000 is the default)
\end{keyword}

\end{frontmatter}

\section{Introduction}
Neuron growth and neurodevelopmental disorders (NDDs) have been some of the most challenging problems to understand due to the intricate physiology of the human brain. 
Although two-dimensional (2D) cell cultures have long been an accessible approach for studying neurodevelopmental and disease processes, they cannot capture the complexity of living neuron cells. 
Recently, much research has shifted focus toward three-dimensional (3D) neuronal cell culture platforms~\cite{frimat2015advances,costamagna2019ipscs}, along with the development of human brain organoids derived from pluripotent stem cells~\cite{amin2018building,lee20173d}. 
These 3D approaches capture essential aspects of brain structures and have shown great promise in studying disorders~\cite{jorfi2018three}. 
However, while 3D cultures represent a significant step forward, they still face the same cost and time limitations and can not efficiently capture the dynamic interactions among neurons ~\cite{hogberg2013toward}. 
To empower these experimental advances, computational frameworks like CX3D~\cite{zubler2009framework} offer the capacity to simulate realistic neurite networks in 3D spaces. 
Yet, accurately analyzing and understanding the neurodevelopmental process and neurological disorders requires an in-depth understanding of different biophysics processes.
This calls for more comprehensive and accurate 3D modeling approaches. 
By continuously improving 3D computational models for neuron growth and neurological disorders, researchers can more accurately simulate and predict physiological processes. 
This enhances our insight into disease mechanisms and informs therapeutic interventions.

Isogeometric analysis-based (IGA) phase field model leveraging truncated T-splines to simulate the multi-stage process of neuron growth has been introduced to investigate NDDs~\cite{qian2025neurodevelopmental}.
This model integrates the dynamics of tubulin concentration and synaptogenesis, providing insights into their functional roles in the neurodevelopmental process. 
However, neurons naturally form complex 3D neurite networks with sophisticated morphologies~\cite{spijkers2021directional}, and relying on 2D analysis significantly limits the potential of the model to understand the full complexity of neurite morphologies~\cite{liao2023ganglia}.
Capturing these features numerically often requires refined numerical techniques for accurate representation and extensive mesh refinements, which lead to excessive degrees of freedom (DOFs) and heightened computational costs.
This challenge underscores the necessity of an efficient 3D geometry representation that maintains smoothness, supports local refinements, and remains analysis-suitable for the phase field method. 
To address this limitation, we extend our model to 3D by using truncated hierarchical B-splines (THB-splines) as the foundation for the IGA implementation for the phase field method. 
This approach enables the simulation of intricate 3D neurite outgrowth with unprecedented detail and accuracy. 
The adoption of THB-splines addresses this need by enabling localized mesh refinements while minimizing computational overhead, offering a scalable solution for modeling the complex dynamics of 3D neuron growth.

Since simulating 3D neurite growth and NDDs involves solving highly non-linear, coupled equations that require extensive computational resources~\cite{qian2023biomimetic}, these simulations require local refinements to accurately resolve the intricate morphometric features and spatial complexities of neurite structures. 
Local refinement of splines has been extensively studied through methodologies such as HB-splines, THB-splines, T-splines~\cite{zhang2018geometric, wei2017truncated, sederberg2003t}, LR-splines~\cite{dokken2013polynomial, bressan2013some}, PHT-splines~\cite{deng2008polynomial}, and truncated hierarchical Catmull-Clark subdivision~\cite{wei2015truncated, wei2016extended, wei2017THTS}.
Further improvements include the analysis-suitable T-splines (ASTS) framework~\cite{wei2022analysis} and joint segmentation methods~\cite{pawar2019joint}, which help refine spline meshes for accurate numerical simulations.
HB-splines~\cite{forsey1988hierarchical} introduced local refinement by hierarchically overlapping coarser and finer B-splines, but their implementation suffers from redundancy due to overlapping functions. 
THB-splines~\cite{giannelli2012thb}, an advancement over HB-splines, incorporate a truncation mechanism to enforce the partition of unity, reducing overlaps and enabling efficient refinement. 
This refinement strategy is well-suited for the challenges of simulating 3D neurite growth. 
By supporting multi-level local adjustments, THB-splines achieve high precision in representing complex neurite morphologies on coarse meshes. 
This capability is essential for capturing detailed, nonlinear behaviors of neurite outgrowth and providing a scalable solution for modeling 3D neurite structures.

In this paper, we introduce a 3D extension of our previous 2D NDDs model by employing THB-splines for flexibility and robust local refinement capabilities. 
Leveraging these advantages, the model overcomes the dimensional constraints of its 2D predecessor, enabling high-fidelity representations of neurite structures in 3D. 
This significantly improves computational efficiency and simulation accuracy, establishing the 3D model as a robust framework for studying 3D neuron growth and NDDs. 
Moreover, moving to 3D lays the foundation for a more in-depth exploration of NDDs, offering critical insights that can ultimately guide therapeutic treatment planning.
The main contributions of this paper include:
\begin{itemize}
    \item Development of a 3D IGA-based phase field computational model for neurodevelopmental process, bridging the gap between biological conceptual modeling and the full spatial complexity of 3D neuron morphologies;
    \item Implementation of THB-splines with multi-level local refinement, enabling efficient computation of complex phase field evolutions in 3D while preserving high accuracy at neuron cell and neurite boundary interfaces;
    \item Introduction of specialized algorithms including unique 3D neuron identification coupled with refined 3D tip detection, and 3D dynamic domain expansion for efficient simulation, which together enable biomimetic modeling of outgrowth dynamics and the analysis of interactions in multi-neuron simulations; and
    \item Conducted a comprehensive NDDs study leveraging the 3D neuron growth model, revealing intricate neurite growth behaviors, highlighting morphological complexities, and paving the way for more effective therapeutic interventions. 
\end{itemize}

The remainder of this paper is organized as follows.
Section~\ref{sec:overview} gives an overview of the 3D phase field neuron growth model. 
Section~\ref{sec:THB_review} provides a review of truncated hierarchical B-splines and multi-level local refinements.
Section~\ref{sec:3D Neuron Growth Model} elaborates on constructing and adapting our 3D phase field neuron growth model.
Section~\ref{sec :3D implementation} walks through the 3D implementation specifics of our model.
Section~\ref{sec:3D_results} presents the results of the 3D neuron growth simulations
% , including detailed NDDs case studies 
that showcase the capabilities of the model.
Finally, Section~\ref{sec:conclusion} concludes our findings and discusses directions for future research.

\section{Overview of 3D Neuron Growth Model}
\label{sec:overview}

\begin{figure*}[ht]
    \centering
    \includegraphics[width=\textwidth]{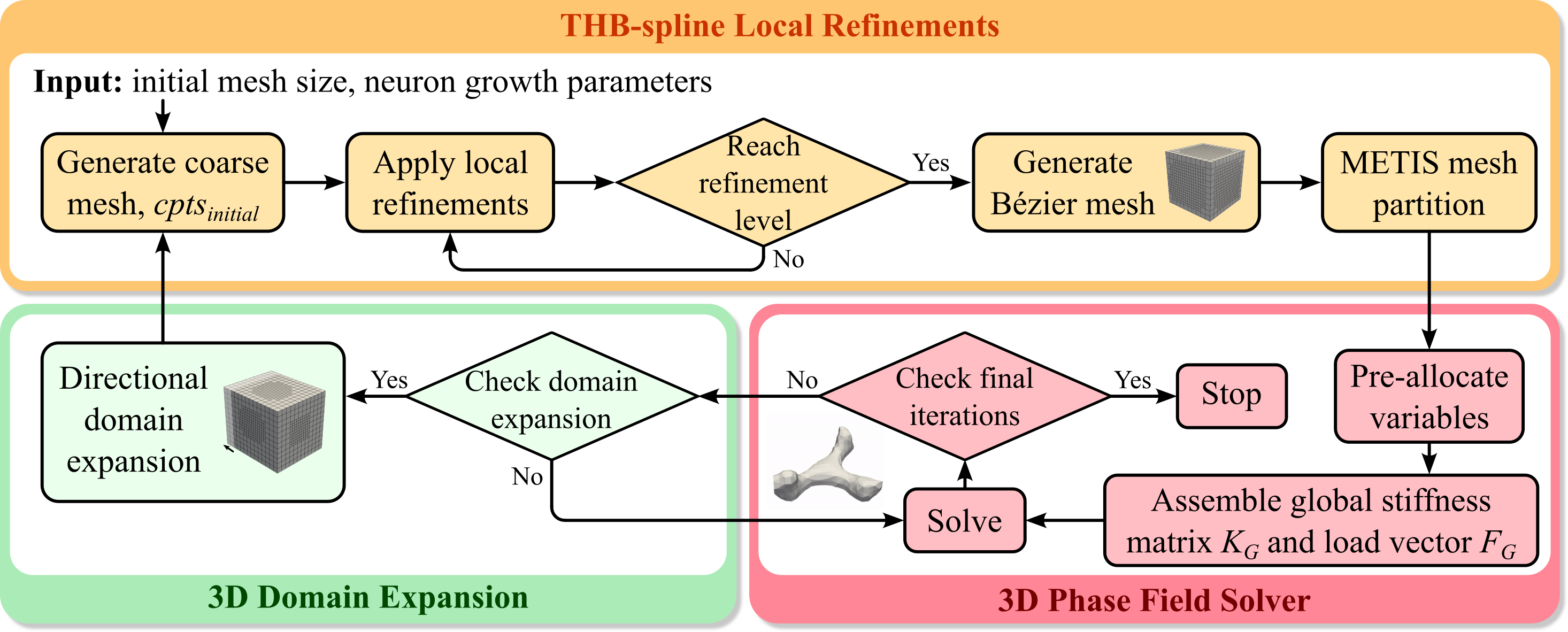}
    \caption[Overview of 3D neuron growth computational model.]
    {Overview of 3D neuron growth computational model. 
    (Orange Module) 3D model preprocessing and parallelization pipeline that handles local refinements and mesh preparations.
    (Green Module) 3D mesh domain expansion module that expands the domain as the neurites approach the boundary.
    (Red) 3D phase field solver that solves the phase field model to simulate neurite morphological transformations.
    }
    \label{fig:overview}
\end{figure*} 

In this paper, we develop a comprehensive 3D neuron growth model to address the limitations of existing 2D frameworks. 
The previous NDDs model~\cite{qian2025neurodevelopmental}, constructed using truncated T-splines with local refinements, enables efficient simulations of neurite growth and deterioration in 2D, but can not capture the intricate 3D morphological transformations of neuron structures, limiting its ability to analyze the spatial complexities often observed in experimental datasets and constraining its potential applications, such as ML-based neurite deterioration prediction~\cite{qian2025nddml}.

To overcome the challenges of modeling 3D neuron growth, the framework depicted in Figure~\ref{fig:overview} integrates trivariate THB-splines with localized refinements. 
THB-splines enable precise, adaptive mesh refinement in regions of interest, ensuring enhanced resolution where needed while maintaining computational efficiency.
This approach effectively captures the intricate morphometric details of 3D neurites and simulates their dynamic growth processes with high fidelity.
The workflow consists of three main components:
\begin{itemize}
    \item \textit{Trivariate THB-spline with local refinements.}
    The neuron growth model initializes a coarse mesh and progressively applies local refinements until the desired resolution is reached.
    A Bézier mesh is then generated and partitioned using METIS for optimal parallel computation.
    \item \textit{3D domain expansion.}
    Dynamic domain expansion ensures that the simulation domain adapts to the directional growth of neurites by evaluating boundary conditions and expanding the mesh as needed.
    A KD-tree-based interpolation algorithm efficiently transfers variables between old and newly refined meshes, preserving data continuity.
    \item \textit{3D IGA phase field solver.}
    Finally, the phase field solver iteratively solves the governing equations, leveraging the refined mesh to accurately represent the evolving neurite structures in 3D.
    The model checks for final iterations, and upon convergence, the simulation stops.
\end{itemize}
This comprehensive 3D framework advances the study of neurodevelopmental processes, offering a new approach for simulating and analyzing 3D neuron growth and its underlying mechanisms.

% To overcome these challenges, we extend the NDDs model to 3D by leveraging THB-splines with local refinements (Figure~\ref{fig:overview}). 
% THB-splines enable localized mesh refinement in regions of interest, enhancing the resolution where needed while maintaining computational efficiency. 
% This capability allows the model to accurately represent the detailed morphometric features of 3D neurites and effectively simulate their dynamic growth processes.
% The 3D neuron growth model, developed through integrating THB-splines and reformulating phase field governing equations, represents a significant advancement in the study of neurodevelopmental processes. 
% It provides a robust and efficient tool for simulating neuron growth and deterioration. 
% This enables a deeper understanding of NDDs and paves the way for future applications and therapeutic developments.

\section{Review of Truncated Hierarchical B-splines and Multi-level Local Refinements}
\label{sec:THB_review}

\begin{figure*}[ht]
    \centering
    \includegraphics[width=0.95\textwidth]{
        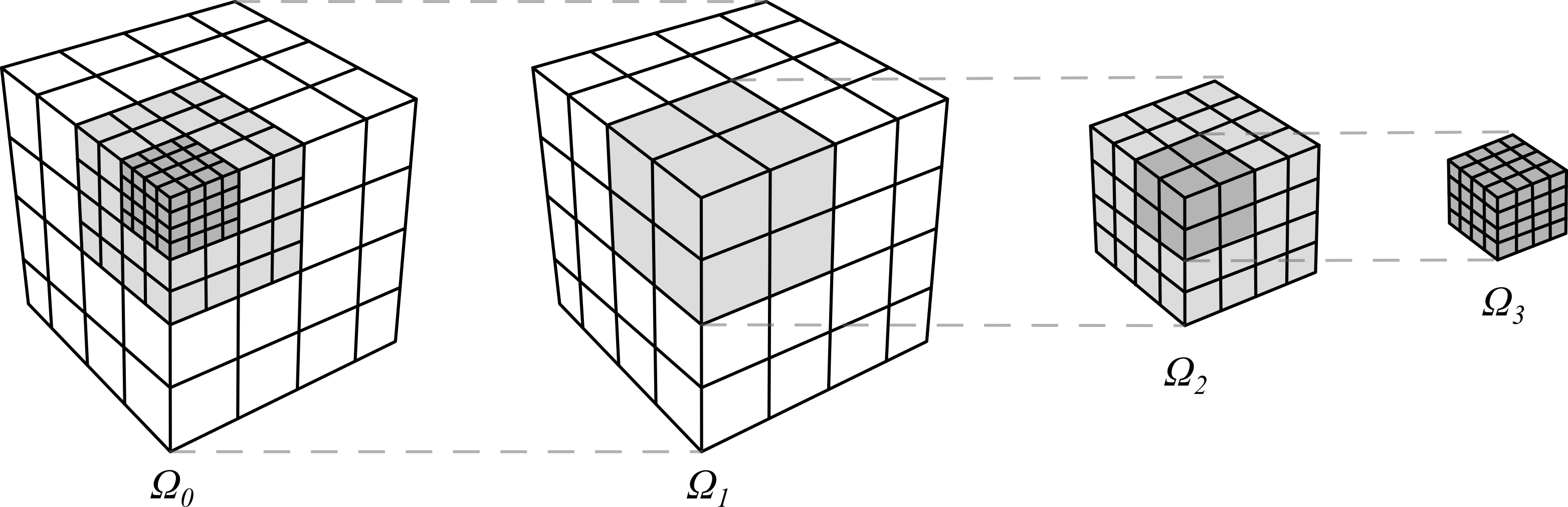
    }
    \caption[3D nested domains for constructing the THB-splines hierarchy.]
    {3D nested domains for constructing the THB-splines hierarchy based on $\Omega_l \supset \Omega_{l+1}$ for $l = 0, 1, ...$. Gray elements are selected for local refinements.
    In the context of $\phi$ interface-based local refinements for neuron growth, we use $\phi$ value to select refinement region.}
    \label{fig:THB_localrefinement}
\end{figure*} 

IGA has significantly advanced computational mechanics by integrating computer-aided design (CAD) and finite element analysis (FEA), particularly through hierarchical and locally refined splines~\cite{vuong2011hierarchical}. 
Among these splines, THB-splines have proven to be effective for efficiently solving phase field problems~\cite{pawar2018dthb3d}. 
They are especially well-suited to capturing the intricate morphologies of 3D neurite growth. 
Since the phase field method uses a diffuse interface to approximate sharp boundaries, it relies on high-resolution meshes around moving interfaces to ensure accurate convergence~\cite{pawar2016adaptive}. 
% Concurrently, advancements in spline technology, such as truncated Catmull-Clark splines~\cite{wei2015truncated}, extended Catmull-Clark splines~\cite{wei2016extended}, and truncated T-splines~\cite{wei2017truncated}, continue to enhance IGA. 
% Despite success with 2D phase field neuron growth model~\cite{qian2025neurodevelopmental}, truncated T-splines still face practical challenges despite their conceptual maturity. 
To solve the 3D neuron growth problem, we exploit the strengths of THB-splines by integrating them with phase field methods. 
We efficiently resolve complex morphologies through multi-resolution grids and adaptive local refinement. 
Because phase field formulations approximate sharp interfaces with diffuse transitions, localized refinement around these moving boundaries substantially boosts computational accuracy and efficiency. 
Specifically, we adopt a multi-resolution grid approach in which THB-splines are refined at each grid resolution near the evolving phase field boundary. 
This approach focuses the computational resources on the most needed regions, enabling high-fidelity simulations of 3D neurite growth while maintaining tractable computational costs.

THB-splines are an extension of HB-splines designed to improve local refinement capabilities~\cite{giannelli2012thb}. 
The construction of HB-splines involves defining a hierarchy of $N$ levels of nested spline spaces $S^0 \subset S^1 \subset \cdots \subset S^N$, over corresponding nested domains $\Omega_N \subset \Omega_{N-1} \subset \cdots \subset \Omega_0$ (Figure~\ref{fig:THB_localrefinement}).
In the context of 3D THB-splines, trivariate B-spline basis functions $B(u, v, w)$ are constructed from the tensor product of three univariate basis functions defined on open knot vectors $U = \{u_1, u_2, \ldots, u_{n_1+p+1}\}$, $V = \{v_1, v_2, \ldots, v_{n_2+p+1}\}$, and $W = \{w_1, w_2, \ldots, w_{n_3+p+1}\}$. 
Here, $n_1$, $n_2$, and $n_3$ represent the number of univariate B-spline basis functions in each parametric direction, and $p$ is the spline order.
We choose $p=3$ for all the case studies in this paper. 
The local support of each basis function $B_{i,j,k}(u, v, w)$ is defined as 
\begin{gather}
    supp(B_{i,j,k}) = [u_i, u_{i+p+1}] \times [v_j, v_{j+p+1}] \times [w_k, w_{k+p+1}],
\end{gather}
where $B_{i,j,k}(u, v, w) = N_i(u) N_j(v) N_k(w)$ and $N_i(u)$, $N_j(v)$, and $N_k(w)$ are the univariate basis functions along the $u$, $v$ and $w$ directions, respectively.
At any refinement level $l$, each trivariate basis function $B^l(u, v, w)$ in the space $S^l$ can be expressed as a linear combination of its children trivariate basis functions $B^{l+1}_{i,j,k}(u, v, w)$:
\begin{equation}
    % B^l(u, v, w) = \sum_{m=1}^{N_c} c_m B^{l+1}_m(u, v, w),\\
    B^l(u, v, w) = \sum_{i=1}^{N_i} \sum_{j=1}^{N_j} \sum_{k=1}^{N_k} c_{i,j,k} B^{l+1}_{i,j,k}(u, v, w),
    \label{eqn: BL}
\end{equation}
where $N_i$, $N_j$, and $N_k$ are the number of children basis functions in each dimension. 
We choose degree-$p$ for each parametric direction, therefore $N_i = N_j = N_k = p+2$. 
The support of each child basis function $chd(B^l)$ or $B_{i,j,k}^{l+1}$ is nested within the support of its parent $B^l$.
The coefficients $c_{i,j,k}$ are determined using Oslo's knot insertion algorithm~\cite{cohen1980discrete}.

To maintain the partition of unity property in THB-splines during refinement, basis functions at the coarser level with duplicated contributions overlapping across levels are truncated~\cite{wei2017THTS, pawar2018dthb3d}. 
Here we use univariate basis functions to explain the truncation mechanism. During the refinement process from a coarser level $l$ to a finer level $l+1$, overlapping level-$l$ functions are selected and become \textit{passive}, while their corresponding \textit{children} basis functions at Level $l+1$ are introduced and become \textit{active}, forming active set $\mathcal{B}_{a}^{l+1}$. 
Level-$l$ basis functions not selected for refinement remain \textit{active}, staying as part of set $\mathcal{B}_{a}^{l}$. 
Truncation resolves overlaps between the active functions from both levels by modifying the specific subset of active Level-$l$ functions $\mathcal{B}_{t}^{l}$, whose support overlaps with the newly active Level-$(l+1)$ functions. 
This set $\mathcal{B}_{t}^{l}$ contains the active Level-$l$ functions ($B_{i}^{l} \in \mathcal{B}_{a}^{l}$) that possess one or more active children at level $l+1$ ($B_{j}^{l+1} \in \mathcal{B}_{a}^{l+1}$):
% $\mathcal{B}_{t}^{l}=\{B_{i}^{l}: chdB_{i}^{l} \cap \mathcal{B}_{a}^{l+1} \neq \emptyset \}$, 
\begin{equation}
    \mathcal{B}_{t}^{l}=\{B_{i}^{l} \in \mathcal{B}_{a}^{l}: \exists B_{j}^{l+1} \in chdB_{i}^{l} \text{ s.t. } B_{j}^{l+1} \in \mathcal{B}_{a}^{l+1} \}.
\end{equation}
The truncation procedure leverages the aforementioned refinability property (Eqn~\ref{eqn: BL}). 
Then, truncation is performed for each $B_{i}^{l} \in \mathcal{B}_{t}^{l}$ by discarding the contributions from its \textit{active} children:
\begin{equation}
trunB_{i}^{l} =  \sum_{\substack{B_{j}^{l+1} \in chdB_{i}^{l} \wedge B_{j}^{l+1} \notin \mathcal{B}_{a}^{l+1}}} c_{j}B_{j}^{l+1}, \quad \forall B_{i}^{l} \in \mathcal{B}_{t}^{l}.\\
\label{eqn: truncation THB-splines} 
\end{equation}
This ensures that the resulting truncated basis functions, together with the other active functions, maintain the partition of unity property~\cite{wei2017THTS, pawar2018dthb3d}. 
The complete THB-spline basis set for this step is then formed by the union of the non-truncated active Level-$l$ functions ($\mathcal{B}_{a}^{l} \setminus \mathcal{B}_{t}^{l}$), the set of truncated Level-$l$ functions ($trun\mathcal{B}_{t}^{l} = \{ trunB_{i}^{l} | B_{i}^{l} \in \mathcal{B}_{t}^{l} \}$), and the active Level-$(l+1)$ functions ($\mathcal{B}_{a}^{l+1}$). 
The recursive application of this process extends to multiple levels of local refinements.

In summary, when a basis function at level $l$ is selected for refinement, it is marked as passive during local refinement. 
This passive basis function is then replaced by the summation of its active children basis functions using Eqn.~\ref{eqn: BL}.
The active basis functions at refinement level $l$ having partial support in $\Omega_{l+1}$ are truncated using Eqn.~\ref{eqn: truncation THB-splines}.
Then, active basis function from both levels $l$ and $l+1$ are collected as:
\begin{equation}
\mathcal{B}^{THB}(u, v, w) = \mathcal{B}_a^l(u, v, w) \cup \mathcal{B}_a^{l+1}(u, v, w),
\label{eqn: local refinement THB-splines}
\end{equation}
where the union combines the basis functions from the current and the subsequent levels. 
The active basis functions from both levels $l$ and $l+1$ are then utilized to construct the truncated hierarchical basis.
Note that not all the active basis functions at level $l$ need to be truncated, in other words, $\mathcal{B}_t^l \subset \mathcal{B}_a^l$. 
This approach reduces the overlap of basis function support from coarser levels, reducing the intersection with finer-level basis functions and significantly lowering the DOFs needed for complex neurite structures in 3D.

\section{3D Phase Field Neuron Growth Model}
\label{sec:3D Neuron Growth Model}

3D neuron growth models can provide a powerful platform for studying the intricate processes of neurodevelopmental processes, enabling simulations and analysis of neurite outgrowth and connectivity in a realistic setting. 
These models leverage advanced computational techniques to capture the complex interactions between neuron growth factors (NGFs), biophysics processes, and concentrations that shape neuron morphology.  
However, extending the original model to three dimensions requires modifying the governing equations, which were initially formulated and validated for 2D domains~\cite{qian2023biomimetic, qian2022modeling}. 
This section showcases the 3D neuron growth model, focusing on two key areas. 
First, we develop a 3D model that can capture the fundamental characteristics of healthy neuron growth morphology. 
Then, we extend it to study the etiology of NDDs, highlighting the model potential to aid the development of novel therapeutic strategy.

\subsection{Health Neuron Growth Model}

Our previous neuron growth model only accounts for 2D anisotropy~\cite{qian2025neurodevelopmental}.
The main phase field equation must be extended to capture the geometric complexity and anisotropic growth dynamics inherent in 3D space. 
Here, we incorporate 3D anisotropy terms to simulate intricate neurodevelopmental processes, branching, and deterioration with biological fidelity~\cite{ren_controllable_2018, karma1998quantitative}.
To address the diverse neuron morphological characteristics, parameters for the phase field NDDs model are derived from existing literatures~\cite{takaki_phase_field_2015, nella2022bridging, diehl2016efficient} and then empirical fine-tuned to ensure biomimetic growth behaviors. 
Parameters can be adjusted further to capture biomimetic simulations of specific neuron types~\cite{qian2023biomimetic}.
The modified 3D phase field model is formulated as the following:
\begin{gather}
    \frac{\partial \phi}{\partial t} = M_{\phi} \biggl[ \nabla \cdot \left( a(\Psi)^2 \nabla \phi ) + \frac{\partial}{\partial x} ( a(\Psi) \frac{\partial a(\Psi)}{\partial (\frac{\partial \phi}{\partial x})} (\nabla \phi)^2 )
    + \frac{\partial}{\partial y} ( a(\Psi) \frac{\partial a(\Psi)}{\partial (\frac{\partial \phi}{\partial y})} (\nabla \phi)^2 ) 
    + \frac{\partial}{\partial z} ( a(\Psi) \frac{\partial a(\Psi)}{\partial (\frac{\partial \phi}{\partial z})} (\nabla \phi)^2 \right) 
    \notag \\ 
    + \phi(1-\phi)(\phi - 0.5 + F_{driv} + 6H |\bigtriangledown \theta|) \biggl], \label{eqn: phase field equation 3D}
\end{gather} 
where $\phi$ is the evolving phase field variable representing the neurite outgrowth, $M_{\phi}$ is the mobility coefficient set as 10, $a(\Psi)$ is the anisotropy coefficient, $F_{driv}$ is the driving force for growth, $H$ is a constant value set as 0.007, and $\theta$ is the orientation term randomized between 0 and 1. 
The 3D anisotropy $a(\Psi)$ is modeled using:
\begin{gather}
    a(\Psi) = \bar{a} (1 - 3\xi) \left\{ 1 + \frac{4\xi}{1 - 3\xi} \frac{\left( \frac{\partial \Psi}{\partial x} \right)^4 + \left( \frac{\partial \Psi}{\partial y} \right)^4 + \left( \frac{\partial \Psi}{\partial z} \right)^4}{|\nabla \phi|^4} \right\},
    \label{eqn:anisotropy_function}
\end{gather}
where $a(\Psi)$ is the 3D anisotropy coefficient for the anisotropy gradient~\cite{karma1998quantitative}.
$\bar{a}$ is the average anisotropy magnitude set as 0.50.
$\xi$ is the anisotropy strength set as 0.01. 
This extension enables a more comprehensive and realistic simulation of neuron growth processes, capturing the complex 3D anisotropy. 

The driving force equation incorporates the effect of tubulin into the phase field~\cite{qian2025neurodevelopmental}:
\begin{align}
F_{driv} = \frac{\alpha}{\pi} \arctan\Bigl[
H_\epsilon\Bigl(\frac{dL}{dt}\Bigr) \gamma \bigtriangleup c_{neur} \Bigr],
\label{eqn: driving_force_equation}
\end{align}
where $\frac{\alpha}{\pi} $ is a scaling constant, $H_\epsilon$ is the Heaviside step function, and $\gamma$ is the interfacial energy constant set as 10. 

The intracellular tubulin equation, which governs tubulin transport during neurite elongation, is formulated as:
\begin{equation} 
    \frac{\partial (\phi \,c_{tubu})}{\partial t} 
    = \delta_t \nabla\cdot (\phi \, \nabla c_{tubu}) 
    - \alpha_t \cdot \nabla (\phi \, c_{tubu})
    - \beta_{t} \phi \, c_{tubu} + \epsilon_0 \frac{|\nabla(\phi_0)|^2}{\int |\nabla(\phi_0)|^2 d \Omega} \label{eqn: tubulin_equation} 
\end{equation}
where $c_{tubu}$ represents the tubulin concentration, $\delta_t$ is the diffusion rate set as 4, $\alpha_t$ is the active transport coefficient set as 0.001, $\beta_t$ is the decay coefficient set as 0.001, and $\epsilon_0 \frac{|\nabla(\phi_0)|^2}{\int{|\nabla(\phi_0)|^2} \, d\Omega}$ is the constant production term. 
In this notation, $\phi_0$ is the initial phase field, and $\epsilon_0$ is the production coefficient set as 15.

The competitive tubulin consumption equation captures localized tubulin consumption at neurite tips:
% \scriptsize
\begin{align}
\frac{dL}{dt} = r_{g} c_{tubu} - s_{g},
\label{eqn: dLdt_equation}
\end{align}
% \normalsize
where $\frac{dL}{dt}$ characterizes the dynamic tubulin utilization required for neurite extension~\cite{mclean2004mathematical}, while $r_g$ set as 5 and $s_g$ set as 0.05 denoting the assembly and disassembly rates of tubulin, respectively~\cite{mclean2004continuum, van_ooyen_competition_2001}.

The synaptogenesis equation governs the concentration of synaptogenesis particles such as neurotrophin~\cite{qian2023biomimetic}:
% \scriptsize
\begin{align}
\frac{\partial c_{neur}}{\partial t}
= &D_c \nabla^2 c_{neur}
+ K \frac{\partial \phi}{\partial t},
\label{eqn: synaptogenesis_equation}
\end{align}
% \normalsize
where $c_{neur}$ is the neurotrophin concentration, $D_c$ is the diffusion coefficient set as 3, and $K$ is the latent neurotrophin set as 1.8. 
% The term $k_{p75} c_{neur}$ models degradation due to $p75NTR$ receptor binding, while $k_2 c_{neur}$ serves as a sink for neurotrophin~\cite{nella2022bridging, krewson1996transport}.

This 3D model focuses on adapting the anisotropy terms to capture the spatial variations and directional dependencies inherent in 3D neuronal structures. 
This ensures that complex neurite morphologies and dynamic behaviors are accurately represented.
In this 3D model, the phase field equation (Eqn.~\ref{eqn: phase field equation 3D}) is coupled with the tubulin equation (Eqn.~\ref{eqn: tubulin_equation}) and the neurotrophin equation (Eqn.~\ref{eqn: synaptogenesis_equation}) through the driving force term \(F_{driv}\) (Eqn.~\ref{eqn: driving_force_equation}) and the competitive tubulin consumption term \(\frac{dL}{dt}\) (Eqn.~\ref{eqn: dLdt_equation}). 
This setup provides a unified computational environment for simulating neurite outgrowth in three dimensions. 
A detailed list of model parameters and variables is provided in Table~\ref{Table:3D_neuron_parameters} for clarity.
For an in-depth discussion of the parameters and IGA phase field for neuron growth, readers are referred to our earlier works~\cite{qian2022modeling, qian2023biomimetic}.

\begin{table}[ht]
\caption{Parameters utilized in the 3D phase field neuron growth model.}
\vspace{-0.7cm}
\label{Table:3D_neuron_parameters} % Changed label slightly for uniqueness
\begin{center}
\resizebox{\columnwidth}{!}{\begin{tabular}{c l c|c l c}
\toprule
\textbf{Parameter} & \textbf{Description} & \textbf{Value} &
\textbf{Parameter} & \textbf{Description} & \textbf{Value}\\
\hline
% Row 1
$\phi$ & Phase field variable & - & 
$c_{tubu}$ & Tubulin concentration & - \\ 
% Row 2
$M_{\phi}$ & Mobility coefficient & $10$ & 
$\delta_t$ & Tubulin diffusion rate & $4$ \\
% Row 3
$a(\Psi)$ & Anisotropy coefficient & - & 
$\alpha_t$ & Tubulin active transport coefficient & $0.001$ \\
% Row 4
$H$ & Constant in phase field eq. & $0.007$ & 
$\beta_t$ & Tubulin decay coefficient & $0.001$ \\
% Row 5
$\theta$ & Orientation term & Random [0, 1] & 
$\epsilon_0$ & Tubulin production coefficient & $15$ \\
% Row 6
$\bar{a}$ & Average anisotropy magnitude & $0.50$ & 
$\phi_0$ & Initial phase field variable & - \\
% Row 7
$\xi$ & Anisotropy strength & $0.01$ & 
$\frac{dL}{dt}$ & Competitive tubulin consumption rate & - \\ 
% Row 8
$F_{driv}$ & Driving force term & - & 
$r_g$ & Tubulin assembly rate & $5$ \\ 
% Row 9
$\frac{\alpha}{\pi}$ & Scaling coefficient (driving force) & - & 
$s_g$ & Tubulin disassembly rate & $0.05$ \\ 
% Row 10
$\gamma$ & Interfacial energy constant & $10$ & 
$c_{neur}$ & Neurotrophin concentration & - \\
% Row 11
$H_{\epsilon}$ & Heaviside function & - & 
$D_c$ & Neurotrophin diffusion coefficient & $3$ \\
% Row 12
$k_{p75}$ & Neurotrophin degradation rate ($p75NTR$) & - & 
$K$ & Latent neurotrophin & $1.8$ \\
% Row 13
$k_2$ & Neurotrophin sink rate & - & 
$c_{opti}$ & Optimal neurotrophin concentration & - \\ 
\bottomrule
\\[-1em] % spacing
\multicolumn{6}{l}{Note: Values provided are defaults used for initialization where applicable.}
\end{tabular}}
\end{center}
\vspace{-0.75cm}
\end{table}

\subsection{Neurodevelopmental Disorder Equations}

To model the neurite deterioration associated NDDs, we incorporated neurotrophin concentration into our previous 2D model~\cite{qian2025neurodevelopmental}.  
Neurotrophin diffusion is critical in guiding growth cones and shaping neurite pathways during synaptogenesis~\cite{song1997camp}. 
The regulation of neurotrophin particles, including their degradation, is introduced into the synaptogenesis equation:
% \scriptsize
\begin{align}
\frac{\partial c_{neur}}{\partial t}
= &D_c \nabla^2 c_{neur}
+ (K - k_{p75}c_{neur})\frac{\partial \phi}{\partial t} - k_2 c_{neur},
\label{eqn: NDDs synaptogenesis_equation}
\end{align}
% \normalsize
where $c_{neur}$ is the neurotrophin concentration, $D_c$ is the diffusion coefficient set as 3, and $K$ is the latent neurotrophin set as 1.8. 
The term $k_{p75} c_{neur}$ models degradation due to $p75NTR$ receptor binding, while $k_2 c_{neur}$ serves as a sink term for neurotrophin~\cite{nella2022bridging, krewson1996transport}.
Both $k_{p75}$ and $k_2$ regulate and control the concentration level of $c_{neur}$, which affects the driving force and interface evolution.
$D_c$ is the diffusion rate for $c_{neur}$, responsible for diffusing the $c_{neur}$ needed for neurite survival to different areas of the neurite to drive outgrowth.
Insufficient diffusion caused by inadequate $D_c$ magnitude will lead to phase field interface retraction and, therefore, capturing neurite deterioration.

The driving force equation incorporates the interplay of tubulin and neurotrophin, coupling these effects back into the phase field~\cite{qian2025neurodevelopmental}:
% \scriptsize
\begin{align}
F_{driv} = \frac{\alpha}{\pi} \arctan\Bigl[
H_\epsilon\Bigl(\frac{dL}{dt}\Bigr) \gamma (c_{opti}-c_{neur})\Bigr],
\label{eqn: NDDs driving_force_equation}
\end{align}
% \normalsize
where the term $(c_{opti} - c_{neur})$ reflects the inverse relationship between neurotrophin levels and neuronal survival, introduced via the optimal neurotrophin concentration $c_{opti}$.
The model captures the dynamic interactions between $c_{neur}$ and neurite extension through its influence on the driving force $F_{driv}$ within the double-well potential of the phase field equation (Eqn.~\ref{eqn: phase field equation 3D}). 
When $c_{neur}$ is below a pre-defined $c_{opti}$, $F_{driv}$ exerts a positive influence, pushing the interface out and thus extending neurites. 
Conversely, when $c_{neur}$ exceeds $c_{opti}$, the effect of $F_{driv}$ is reversed, inducing interfacial retraction. 
This bi-directional regulation, modulated by the balance between $c_{opti}$ and $c_{neur}$, enables the simulation of a diverse range of neurite morphological transformations.

% The idea here is that as $c_{neur}$ increases, as long as it is lower than $c_{opti}$, the effect and strength of $F_{driv}$ is positive on the double well term in Eqn.~\ref{eqn: phase field equation 3D}, pushing the interface out.
% However, once $c_{neur}$ rises above $c_{opti}$, the effect of $F_{driv}$ on Eqn.~\ref{eqn: phase field equation 3D} is reversed, leading to interface retraction.
% By adjusting the $c_{opti}$ and $c_{neur}$ balance, the model can simulate different neurite morphological transformation behaviors.

\begin{figure*}[ht]
    \centering
    \includegraphics[width=\textwidth]{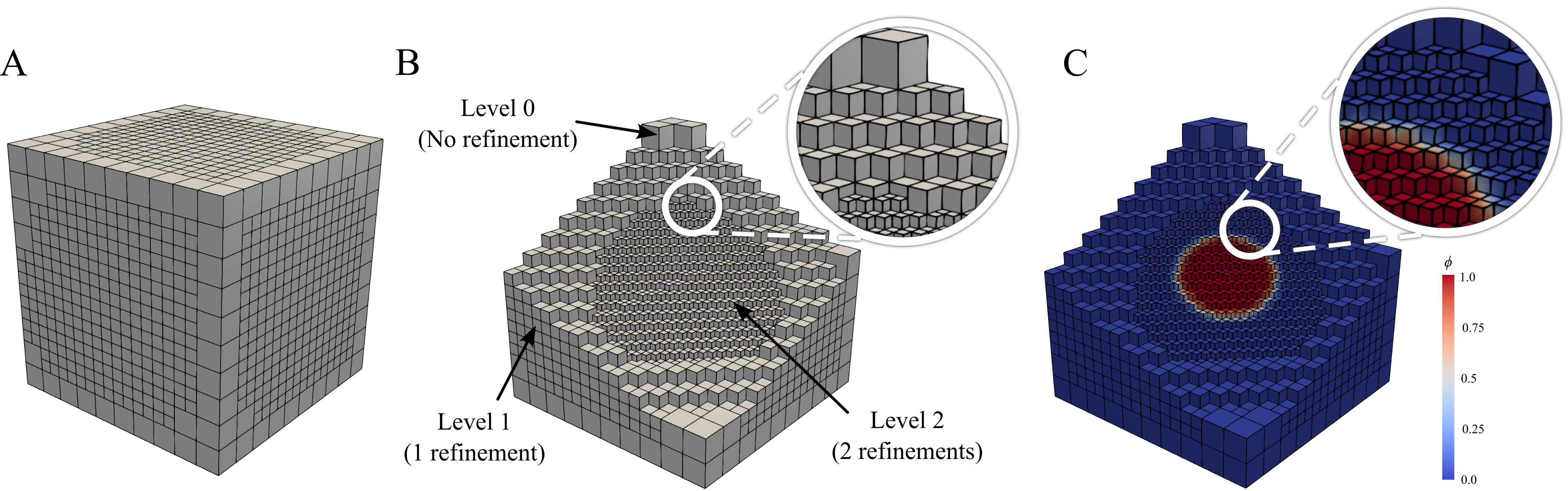}
    \caption[3D phase field-based local refinements on THB-splines.]
    {3D phase field-based local refinements on THB-splines.
    (A) THB-splines with local refinements.
    (B) Cross-section view of three levels of refinement.
    (C) 3D phase field neuron growth variable $\phi$ on locally refined THB-splines.
    }
\label{fig:neuron-thb-spline}
\end{figure*}

\section{3D Numerical Implementation}
\label{sec :3D implementation}

% emphasize 3D contributions over 2D work

% Embed variable values into the text

% parameters values same as 2D paper ~\cite{qian2025neurodevelopmental}

% ideas for later on (2 neuron simulations)

% Start of Paragraph 1

The previous 2D NDDs model~\cite{qian2025neurodevelopmental}, while effective for simulating deterioration using locally refined T-splines, fundamentally lacks the capacity to represent realistic 3D neurite morphology observed experimentally~\cite{bennington2025towards}.  
Extending this work to 3D introduces profound challenges related to geometric representation, algorithmic complexity, and computational scale.  
Our 3D neuron growth model addresses these challenges by adopting a fundamentally different and more sophisticated approach.  
We utilize locally refined THB-splines as the discretization basis, which, while more complex to manage than T-splines, provides essential multi-level local refinement capabilities critical for efficiently handling anisotropic terms inherent in the 3D phase-field equations.  
Furthermore, the 3D model incorporates several newly developed, computationally intensive components presented as key contributions: 
\begin{itemize}
    \item \textit{Adaptive THB-spline Refinement:} Phase-field-driven ($\phi$) local refinement using THB-splines, focusing compute resources on neuron structures and efficiently handling 3D anisotropy~\cite{qian2025neurodevelopmental};
    \item \textit{3D Neuron Identification:} Uniquely labeling distinct neuron instances, preventing self-intersection errors during tip detection and interaction analysis~\cite{qian2022modeling};
    \item \textit{3D Neurite Tip Detection:} Locating active growth tips by analyzing $\phi$ intensity patterns, directing neurite elongation and branching events~\cite{qian2023biomimetic}; and
    \item \textit{Dynamic Domain Expansion:} Adaptively expanding the simulation domain as neurites approach boundaries, enabling efficient simulation of extended growth while avoiding prohibitive pre-allocation costs.
\end{itemize}

The simulation of 3D neuron morphologies, which evolve and expand dynamically, poses significant computational demands. 
Beginning with locally refined THB-spline control meshes with domains of approximating $20\times20\times20$ elements, the control mesh adaptively expands during growth simulation, leading to over six million DOFs in later stages. 
This estimate considers the coupled phase-field, tubulin, and neurotrophin equations~\cite{qian2023biomimetic, qian2025neurodevelopmental} solved using tri-cubic 3D Bézier elements. A KD-tree~\cite{bentley1975multidimensional} based algorithm, using Nanoflann~\cite{blanco2014nanoflann}, handles the necessary interpolation of variables onto this expanding control mesh. 
High-performance computing techniques are required to manage the computational cost. 
The C++ implementation utilizes the PETSc library~\cite{petsc-user-ref} for its parallel solvers, employing Newton-Raphson for the non-linear phase-field problem, implicit Euler for time integration, and preconditioned GMRES~\cite{saad1986gmres} for the linear systems.
Key optimizations, including variable pre-computation and minimizing nonlinear solver operations, were also implemented. MPI~\cite{petscsf2022, gropp1996high} parallelization combined with METIS~\cite{karypis1998fast} for load balancing after refinement ensures scalability, with computations on the Bridges-2 supercomputer~\cite{ecss, xsede} scaling from 128 to 256 threads as the problem size grows. 
To handle the 72-hour wall time limit at the Pittsburgh Supercomputing Center, we develop a checkpoint/restart capability, achieved through SLURM and C++ scripts. 
The combination of adaptive high-order discretization, specialized algorithms, optimizations, scalable parallel computing, and restart functionalities provides an accurate and effective adaptive-resolution framework for these complex, evolving neuron growth simulations.

\subsection{Phase Field-Based Local Refinements for 3D THB-Splines}  
\label{sec:phase field-local_refine}

Directly assembling the 3D phase field model solver requires constructing multiple high-order terms at each Gaussian quadrature point for all elements, leading to a substantial number of DOFs. 
This leads to significant computational costs that increase exponentially as the domain expands.
THB-splines, constructed on structured hexahedral control meshes, integrate seamlessly into IGA, bridging the gap between CAD and numerical analysis~\cite{wei2017THTS}. 
By enabling localized refinements in regions where the phase field variable $\phi$ captures neuron morphological transformations, THB-splines minimize unnecessary DOFs and allow the solver to efficiently handle the complex dynamics of evolving interfaces. 
Extending the principles of local refinement from 2D to 3D, THB-splines enable precise element refinement in 3D. 
This ensures high resolution in regions of interest while maintaining computational efficiency across the entire domain, making the approach particularly effective for simulating intricate 3D phase field interface evolution.

\begin{figure*}[ht]
    \centering
    \includegraphics[width=\textwidth]{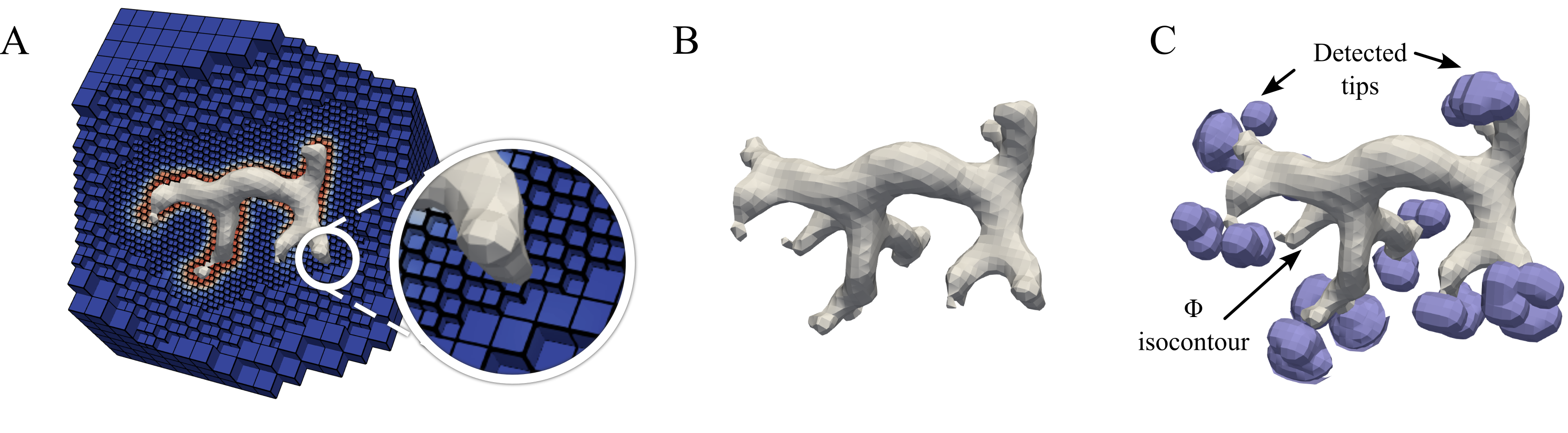}
    \caption[3D tip detection.]
    {3D tip detection visualizations. 
    (A) Clipped view of $\phi$ on locally refined THB-splines with a zoomed-in view of the locally refined mesh.
    (B) Corresponding $\phi$ isocontour surface.
    (C) Detected 3D tips (purple isocontours) on the 3D neurites $\phi$ (grey isocontours).
    % (B) Corresponding clipped view with detected 3D tips (purple isocontours).
    % (C) Three different perspective views of detected 3D tips (purple isocontours) on the 3D neurites $\phi$ (grey isocontours).
    }
\label{fig:3D tip detection}
\end{figure*}

\algrenewcommand\algorithmicrequire{\textbf{Procedure}}
\begin{algorithm}[h]
    \caption{3D Tip Detection on THB-Splines}
    \textbf{Input}: $\phi$ on mesh $\mathcal{M}^{0}$, finer mesh $\mathcal{M}^{fine}$, box size $L$, threshold $\zeta_{tip}$, neuron identification $ID$ \\
    \textbf{Output}: Detected tips $tips$ on ${\mathcal{M}^{0}}$

    \begin{algorithmic}[1]
        \Require{Interpolate $\phi$ from $\mathcal{M}^{0}$ to $\mathcal{M}^{fine}$ \Comment{Interpolate $\phi$}}
        \For{each $p \in \mathcal{M}^{fine}$}
            \State Find $k$-nearest neighbors in $\mathcal{M}^{fine}$
            \State Compute $\phi^{fine}(p) = \frac{\sum_{i=1}^k w_i \phi(p^{0}_{i})}{\sum_{i=1}^k w_i}$, $w_i = \frac{1}{d(p, \mathcal{M}^{fine}_{i}) + \epsilon}$
        \EndFor

        \Require{Compute tip scores $S(p)$} \Comment{Compute scores using $ID(p)$}
        \For{each $q \in \mathcal{M}^{fine}$}
            \If{$ID(q) = ID(p)$} \Comment{Only calculate intensity if $p'$ is part of a neuron}
                 \State $I(p) = I(p) + \phi^{fine}(q)$
            \EndIf
            % Original Line 11 (S(p) calculation) remains unchanged but uses the modified I(p):
            \State $S(p) = \begin{cases}
                            \frac{\phi^{fine}(p)}{I(p)} & \text{if } I(p) > 0 \\
                            0 & \text{otherwise}
                        \end{cases}$
        \EndFor
        % \Require{Compute tip scores $S(p)$. \Comment{Compute scores}}
        % \For{each $p \in \mathcal{M}^{fine}$}
        %     \State Neighbors $q \in \mathcal{B}(p, L)$. \Comment{Neighbors in box}
        %     \State $I(p) = \sum_{q} \phi^{fine}(q)$.
        %     \State $S(p) = \begin{cases}
        %         \frac{\phi^{fine}(p)}{I(p)} & \text{if } I(p) > 0 \\
        %         0 & \text{otherwise}
        %     \end{cases}$
        % \EndFor

        \Require{ Map $S(p)$ back to $\mathcal{M}^{0}$ \Comment{Map scores}}
        \For{each $p \in \mathcal{M}^{0}$}
            \State Interpolate $S(p) = \frac{\sum_{i} w_i S(p_i^{fine})}{\sum_{i} w_i}$  \Comment{Interpolate from $\mathcal{M}^{fine}$}
        \EndFor

        \Require{Threshold tip scores \Comment{Threshold}}
        \For{each $p \ \in \mathcal{M}^{0}$}
            \If{$S(p) > \zeta_{\text{tip}} \cdot \max(S)$} \Comment{Max over all $S$ on $\mathcal{M}^0$}
                \State Mark $p$ as a tip
            \EndIf
        \EndFor

        \State \Return $tips$ on ${\mathcal{M}^{0}}$
    \end{algorithmic}
    \label{alg: 3D_tip_detection}
\end{algorithm}

Our approach begins with a uniform coarse control mesh that defines the initial phase field $\phi$ based on parameters such as the neuron cell center and radius $r_0$. 
Local refinements are then applied to elements where $\phi$ represents the neuron, increasing mesh resolution in areas of interest while maintaining coarser elements in regions far away from the neuron (Figure~\ref{fig:neuron-thb-spline}A\&B). 
After constructing a locally refined truncated hierarchical B-spline, $\phi$ values are interpolated onto the refined mesh using a KD-tree-based method (Figure~\ref{fig:neuron-thb-spline}C)~\cite{bentley1975multidimensional}, ensuring accurate initialization by either directly transferring or interpolating control point values.
This approach optimizes computational efficiency by concentrating refinement on regions where the phase field variable captures the intricate geometries of 3D neuron growth. 
It enables high-fidelity simulations that are critical for advancing neurodevelopmental and NDDs research.

\subsection{3D Neuron Identification}

In simulations involving multiple neuron configurations, differentiating individual neurite morphological evolutions using phase field is essential for performing neuron-specific analysis. 
We develop an technique that utilizes the high-resolution phase field variable $\phi_{fine}$, obtained by interpolating the locally refined phase field onto the fully refined mesh $\mathcal{M}^{fine}$, along with predefined seed points $\{s_k\}$ positioned within each distinct neuron.
The core idea is to first identify all regions corresponding generally to neuronal structures by 
% applying a suitable threshold $\zeta_{phi}$ to the $\phi_{fine}$ values, 
generating a binary mask $\phi_{binary}$ across the fine mesh where $\phi_{binary}=1$ indicates these neuron regions and $\phi_{bindary}=0$ indicates background.. 
Subsequently, we employ a seed-based connectivity analysis, conceptually similar to a flood fill algorithm, starting from initial neuron seed positions. 
Starting from each distinct seed point $seed_k$, this analysis explores the connected region defined by the binary mask ($\phi_{binary}=1$), assigning a unique integer label $k$ to all reachable points associated with that specific seed and storing this result in a label field $ID(p)$.

This procedure generates the label field $ID(p)$ across the fine mesh, where $ID(p)$ indicates which specific neuron the point $p$ belongs to and $ID(p)=0$ denotes surrounding extracellular environments. 
This label field $ID(p)$ is critically leveraged during the subsequent tip detection stage.
Specifically, when computing the local intensity $I(p)$ used to determine the tip score $S(p)$, the summation over neighboring points $q$ is restricted to include only those points that share the same unique neuron label, $ID(q)=ID(p)$.
This crucial step ensures that the computed tip score accurately reflects the local morphology of solely the parent neuron, preventing signal interference from adjacent but distinct neuronal structures, thereby enabling reliable identification of active growth zones essential for guiding outgrowth dynamics.

\subsection{3D Tip Detection}
\label{sec:3D tip detection}

The phase-field variable $\phi$ represents the evolving morphology of the neuron.
It is defined on a locally refined mesh constructed for THB-splines. 
The inherent hierarchical structure of THB-splines enables efficient representation of complex geometries by building spline basis functions across multiple levels of control meshes. 
While this approach offers significant advantages for representing the overall neuron morphology, it poses troubles for tip detections because the control mesh for THB-spline is split across multiple files, and accurate tip detection requires a finer resolution to capture subtle variations in the phase field (Figure~\ref{fig:3D tip detection}A).

To accurately resolve the fine-scale neurite growth features of the phase field, $\phi$, we develop a 3D tip detection algorithm for THB-splines (Algorithm~\ref{alg: 3D_tip_detection}). 
First we interpolate $\phi$ from the THB-spline control mesh, $\mathcal{M}^{0}$, to a finer uniform mesh, $\mathcal{M}^{fine}$, resulting in $\phi_{fine}$. 
The uniform element size in $\mathcal{M}^{fine}$ is equivalent to the element size of the most refined level in $\mathcal{M}^{0}$, ensuring consistent resolution across all elements.
The interpolation is performed using a KD-tree-based method \cite{bentley1975multidimensional}, which efficiently searches for close neighbors and then maps $\phi$ values from $\mathcal{M}^{0}$ to $\mathcal{M}^{fine}$ based on the distance to neighboring control points. 
For a point $p \in \mathcal{M}_{fine}$, the interpolated value $\phi_{fine}(p)$ is computed as:
\begin{gather}
\phi_{fine}(p) = \frac{\sum_{i \in N(p)} \phi(\mathcal{M}_{i}) \cdot w_i}{\sum_{i \in N(p)} w_i}, \quad w_i = \frac{1}{d(p, \mathcal{M}_{i}) + \epsilon}, \notag
\end{gather}
where $N(p)$ denotes the set of neighboring points of $p$ in $\mathcal{M}^{0}$, $d(p, \mathcal{M}_{i})$ is the Euclidean distance between $p$ and $\mathcal{M}_{i}$, and $\epsilon$ is a small constant to prevent division by zero.

    Once $\phi_{fine}$ is defined, we compute a tip score, $S(p)$, for each point $p \in \mathcal{M}^{fine}$. The local intensity $I(p)$ is calculated as:
\[
I(p) = \sum_{q \in Box(p, L)} \Phi(\phi_{fine}(q)),
\]
where $Box(p, L)$ represents the set of points within a cubic bounding box of side length $L$ centered at $p$, and $\Phi(\cdot)$ is a transformation function applied to $\phi_{fine}$ to enhance the neurite regions, implemented as the \texttt{CellBoundary} function.
The tip score $S(p)$ is then defined as:
\[
S(p) = 
\begin{cases} 
\frac{\phi_{fine}(p)}{I(p)}, & \text{if } I(p) > 0, \\ 
0, & \text{otherwise.}
\end{cases}
\]
This approach significantly elevates the local $\phi_{fine}$ values relative to their surroundings~\cite{qian2023biomimetic}, enabling accurate identification of neurite tips.

Subsequently, the computed tip scores are interpolated back onto the original THB-spline control mesh using the same KD-tree-based interpolation method, maintaining consistency between the high-resolution tip detection and the underlying model representation (Figure~\ref{fig:3D tip detection}B). 
Finally, potential neurite tips are identified by applying a dynamic threshold. 
Control points with a tip score exceeding a fraction, $\zeta_{tip}$, of the global maximum tip score are flagged as tips. 
The parameter $\zeta_{tip}$ controls the tip detection and can be adjusted based on experimental data or specific simulation requirements.
This combined approach, leveraging the efficient THB-splines, the KD-tree-based interpolation, and a dynamic thresholding scheme, provides a robust and scalable framework for identifying active regions of neurite growth in our 3D simulations (Figure~\ref{fig:3D tip detection}C). 
Algorithm~\ref{alg: 3D_tip_detection} provides a detailed outline of this process.

\begin{figure*}[ht]
    \centering
    \includegraphics[width=\textwidth]{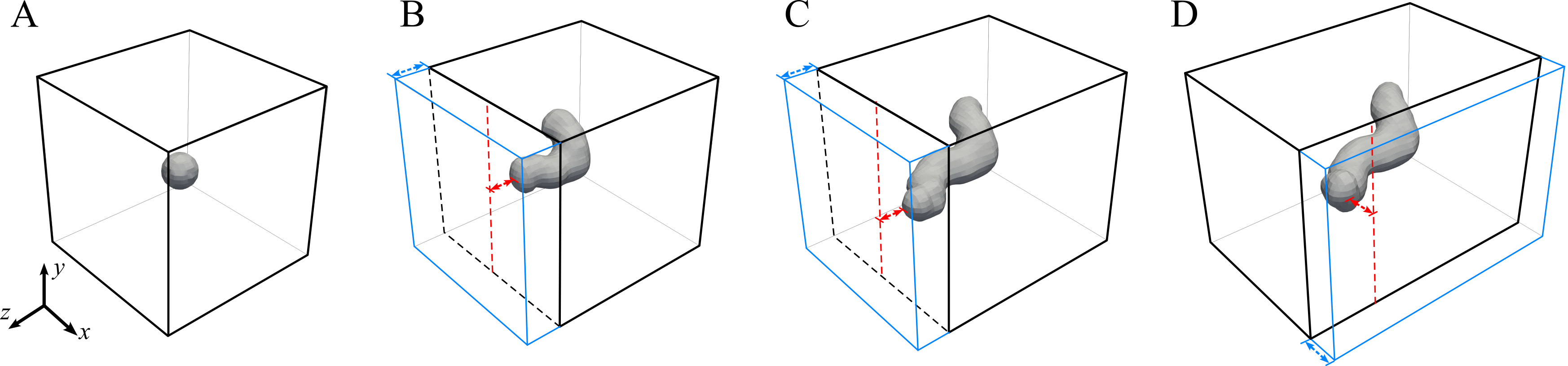}
    \caption[3D neuron growth domain expansion.]
    {3D neuron growth domain expansion. 
    (A) Initial control mesh.
    (B) Domain expands directionally in positive $z$ axis as neurite grows towards the boundary.
    (C) Neurite continues to grow in positive $z$ direction and the domain expands again.
    (D) Neurite turns towards positive $x$ direction and domain expands in positive $x$ direction to accommodate growth.
    }
\label{fig:domain_expansion}
\end{figure*}

\subsection{3D Domain Expansion}
\label{sec:domain_expansion}

Neuron growth initializes near the center of the domain, with neurites gradually extending outward. 
Initializing a large, fixed computational domain for this process is inefficient and computationally expensive, as the phase field variable $\phi$ remains mostly 0 in regions far from the soma. 
To address this inefficiency, neuron outgrowth simulations require adaptive domain adjustments to align with the dynamically expanding neurite elongations. 
A fixed domain leads to unnecessary computational overhead in inactive regions where $\phi = 0$. 
To overcome this challenge, the 3D neuron growth model employs dynamic domain expansion, which allows the control mesh to grow adaptively in response to neurite extension (Figure~\ref{fig:domain_expansion}). 
This approach minimizes unnecessary computational costs while preserving high-resolution accuracy in neuron growth regions.

As shown in Figure~\ref{fig:domain_expansion}, the 3D domain expansion algorithm dynamically expands the simulation domain by first evaluating the phase field variable $\phi$ at elements near the domain boundary. 
If $\phi > 0$ is detected for each boundary element, the corresponding direction is flagged for expansion. 
The domain expands by a size of $1 \times \Delta x$, where $\Delta x$ represents the coarse element size before refinements.
After expansion, local refinement is applied to the control mesh to achieve higher resolution. 
Afterward, each edge of length $\Delta x$ is subdivided into two finer element edges during a single refinement. 
After two refinement steps, each $\Delta x$ edge contains four finer elements for the region with a neuron (Figure~\ref{fig:THB_localrefinement}). 
This expansion and refinement approach balances computational efficiency with the need for high resolution in regions of active neurite growth.
During the expansion, the algorithm generates a new control mesh $\mathcal{M}’$ and shifts its origin to maintain alignment with the phase field variable $\phi$. 
To accurately initialize variables on the expanded mesh, a KD tree-based interpolation method~\cite{bentley1975multidimensional} maps $\phi$ values from the original control points $\mathcal{M}^{0}$ to the expanded control points $\mathcal{M}’$. 
Specifically, if no exact match is found for each $p’ \in \mathcal{M}’$, the nearest neighbors in $\mathcal{M}^{0}$ are identified, and their $\phi$ values are interpolated based on inverse-distance weighting. 
This ensures continuity and accuracy in the simulation variables across the growing domain.

\begin{figure*}[!ht]
    \centering
    \includegraphics[width=\textwidth]{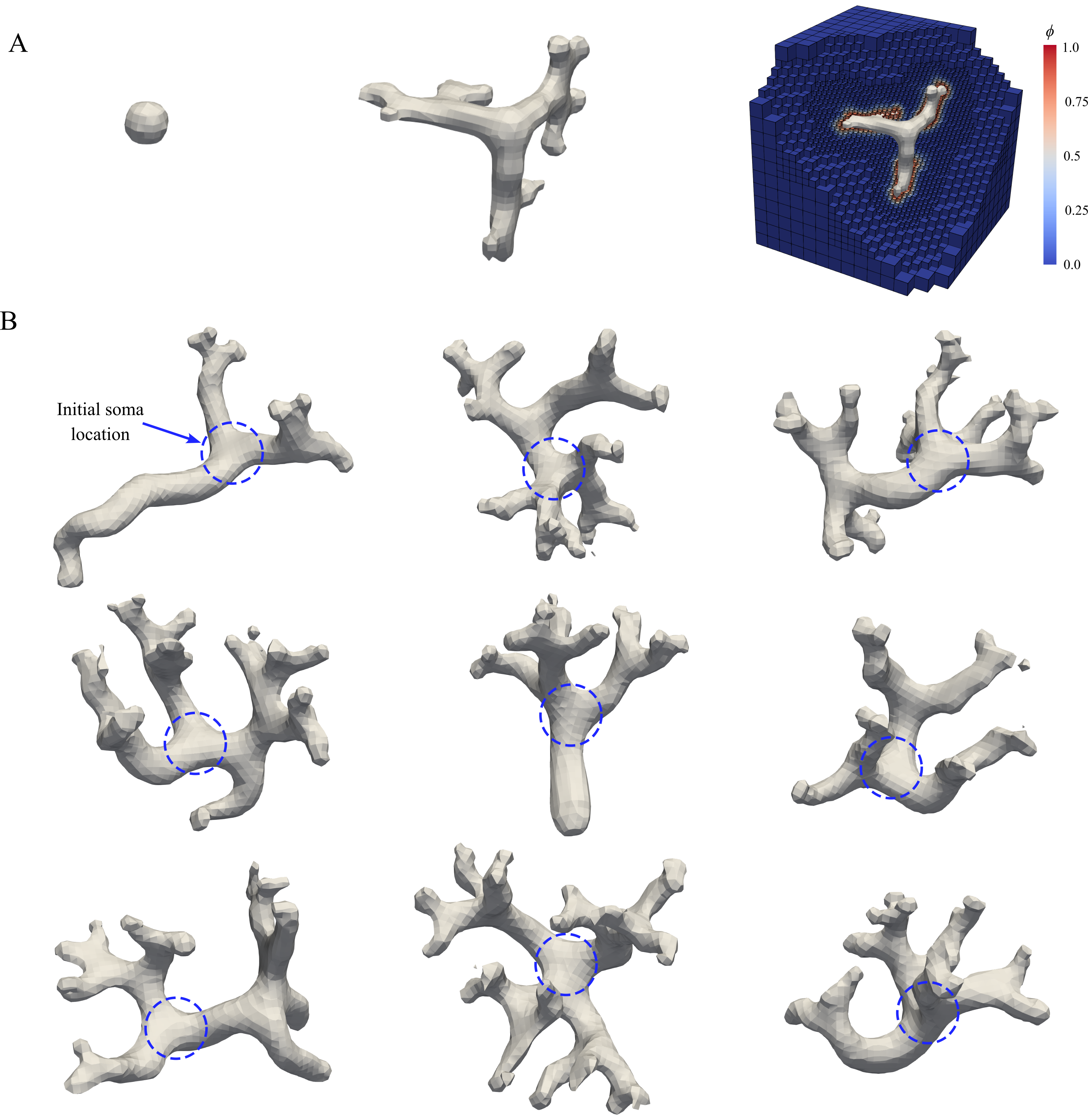}
    \caption[3D neuron growth simulation results of single neuron.]
    {3D neuron growth simulation results of single neuron cases.
    (A) $\phi$ isocontour surface shows the initial neuron soma, branched neurite outgrowth, and corresponding locally refined THB-splines around neurite outgrowth in the 3D domain.
    (B) Single-neuron growth cases with the blue dashed circles representing the initial neuron somas.
    }
    \label{fig:3D_simulations}
\end{figure*} 

\begin{figure*}[!ht]
    \centering
    \includegraphics[width=\textwidth]{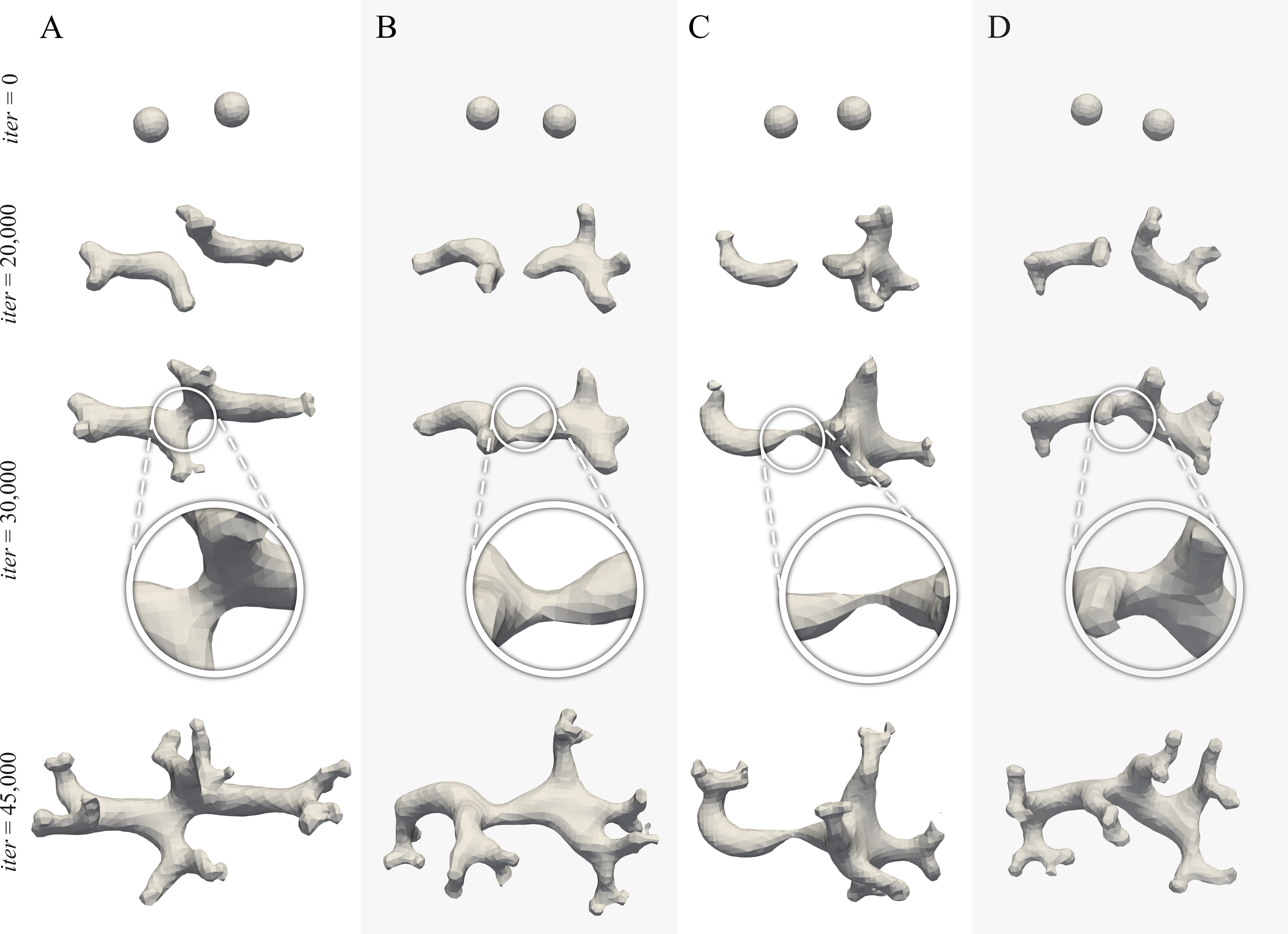}
    \caption[3D neuron growth simulation results of multiple neurons.]
    {3D simulations of multiple neurons with neurite interactions at different iterations. 
    (A-D) Four distinct scenarios, each starting with two neurons (first row, $iter$ = 0). 
    Subsequent columns depict neurite outgrowth (second row, $iter$ = 20,000), the onset of interaction with magnified view of the connection between neurites (third row, $iter$ = 30,000-35,000), and the resulting network structure at later stages (fourth row, $iter$ $\approx$ 44,000-45,000).}
    \label{fig:3D_simulations_multi}
\end{figure*} 

\section{Results}
\label{sec:3D_results}

In this section, we showcase simulation results to evaluate the extended 3D neuron growth model capability.
The simulations effectively capture key aspects of neuron morphology, including neurite elongation, branching, and retraction, demonstrating that the model can adeptly handle complex spatial dynamics. 
By incorporating intracellular transport and neurotrophin signaling, the model accurately captures biologically relevant growth patterns and deterioration behaviors. 
These results validate the robustness of the 3D phase field framework in simulating intricate neurodevelopmental processes and provide critical insights into neurodevelopmental mechanisms, supporting advancements in targeted therapeutic approaches.

\subsection{3D Healthy Neuron Growth}
\label{sec:3D NG simulation results}

Figure~\ref{fig:3D_simulations} shows the results of 3D neuron growth simulations for a single neuron conducted using the phase field-based neuron growth model implemented on locally refined THB-splines. 
The simulations effectively capture the dynamic neurite morphological transformation.
Figure~\ref{fig:3D_simulations}A shows the iso-contour of the phase field variable $\phi$, representing the initial neuron soma and the neurite outgrowth in 3D. 
The locally refined THB-spline mesh highlights the refinement near neuron boundaries, concentrating computational resources on regions with active growth while maintaining coarser meshes in static areas to enhance efficiency.
Figures~\ref{fig:3D_simulations}B illustrates several single-neuron growth cases, demonstrating the model can simulate diverse neuron morphologies. 
Figures~\ref{fig:3D_simulations_multi} shows four additional multiple neuron configurations (2 neurons), showcasing that the model is capable of simulating complex multi-neuron growth and capturing neurite interactions among different neurons.
These cases highlight the capability of our 3D neuron growth model to simulate complex 3D features, such as intricate branching patterns and variations in neurite thickness. 
The adaptive local refinement ensures fine-scale features are resolved while minimizing computational overhead, achieving high accuracy efficiently.

These 3D simulations provide insights into neurodevelopmental processes, including branching, growth trajectories, and thickness variations, which are challenging to capture in 2D models. 
Integrating multi-level local refinement enables accurate representation of dynamic phase field interfaces without excessive computational costs, ensuring the model remains scalable for more complex simulations. 
These results validate the robustness of the 3D neuron growth model and its ability to simulate biomimetic neuron growth behaviors. 
The model provides a solution to bridge computational simulations with experimental observations, enabling more detailed studies of neurodevelopmental processes and neurite network formation. 
This framework provides a foundation for further exploration of larger-scale neurite networks and investigations into NDDs, aiding the development of targeted treatment planning.

\begin{figure*}[htbp]
    \centering
    \includegraphics[width=\textwidth]{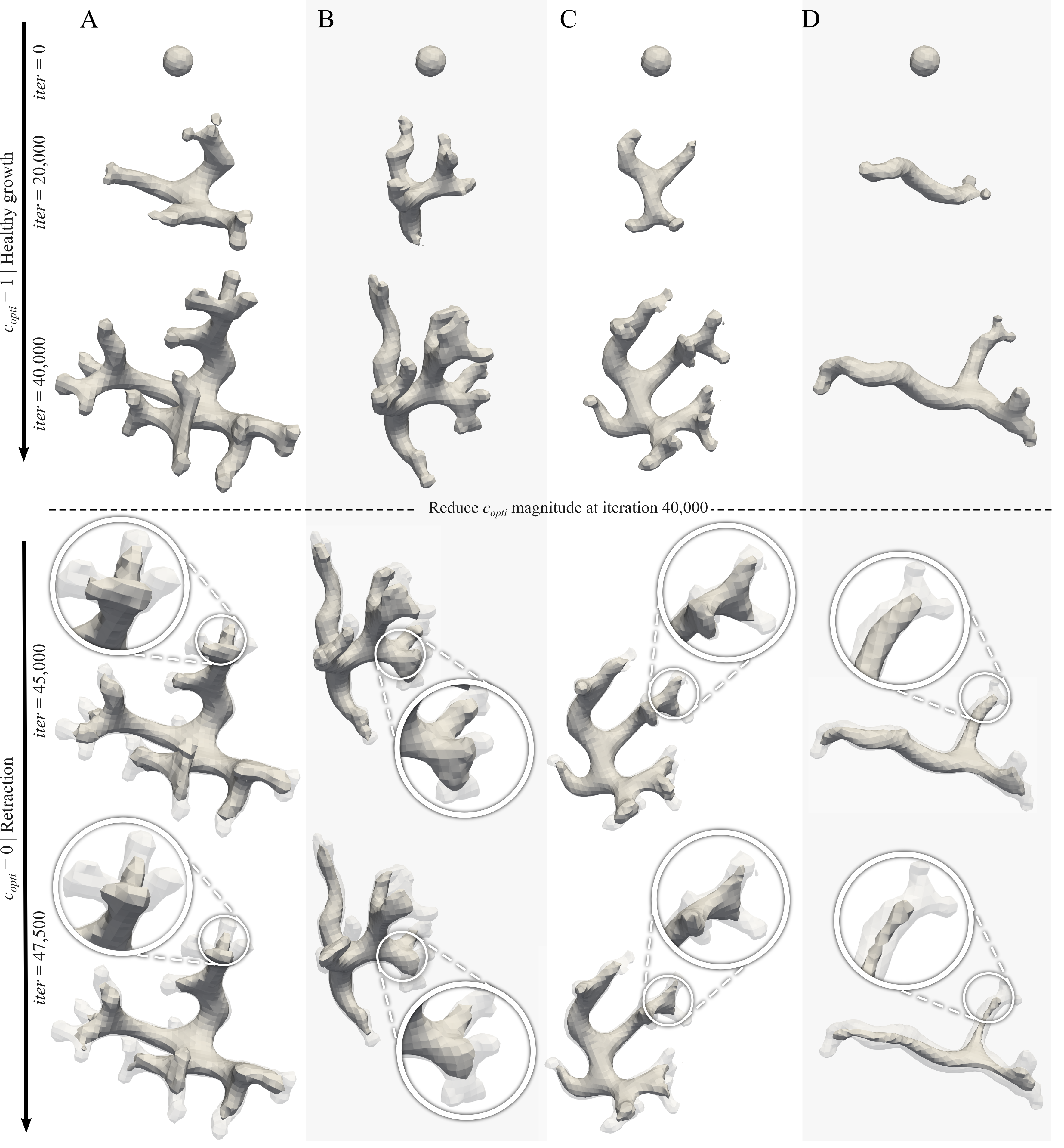} 
    \caption[3D NDD simulation results.]
    {Simulation results illustrating neurite morphological dynamics under changing neurotrophin conditions simulating NDDs. 
    Four distinct simulation cases (A-D) show initial healthy neurite outgrowth (top panels, corresponding to $c_{opti}=1$) followed by subsequent visualization using transparent healthy neurons and zoom-in pictures upon reducing the optimal neurotrophin concentration (bottom panels, $c_{opti}=0$).} 
    \label{fig:3D_NDD}
\end{figure*}

\subsection{3D Neurodevelopmental Disorders Study}
\label{sec:3D NDDs results}

Building upon the foundation of healthy neuron growth simulations, we extend our investigation to explore the onset and progression of NDDs by modulating key biophysical parameters. 
Specifically, we simulate conditions representing neurotrophin deficit, a factor implicated in some NDDs, by altering the optimal neurotrophin concentration parameter, $c_{opti}$. 
As detailed previously~\cite{qian2025neurodevelopmental, piontek1999neurotrophins}, this parameter influences the driving force for neurite extension or retraction. To demonstrate the model capability, simulations are initiated under healthy growth conditions ($c_{opti}=1$) allowing neurites to extend and branch.
Then, $c_{opti}$ is set to $0$ to induce retraction dynamics characteristic of deterioration processes. 

Figure~\ref{fig:3D_NDD} illustrates this process across four distinct simulations (A-D), showcasing healthy neurite outgrowth with $c_{opti}=1$ (top panels) followed by the subsequent retraction behaviors upon setting $c_{opti}=0$ (bottom panels). 
These results highlight the model's capability of capturing intricate and evolving morphological transformations, including atrophy and structural collapse associated with NDDs purely through the modification of the neurotrophin interaction term. 
Leveraging our 3D phase field neuron growth model on locally refined THB-splines enables the investigation of how these factors impact neurite outgrowth and disorder-related morphological transformations.
The precise tracking of complex 3D geometries supported by THB-splines offers a computational tool to gain insights into factors contributing to NDDs, potentially assisting in the development of targeted therapeutic strategies by simulating their effects on neuronal morphology.

\section{Conclusion and Future Work}
\label{sec:conclusion}

Integrating IGA and phase field methods in 3D neuron growth modeling enables the capturing of intricate spatial complexity of neurite morphological structures. 
Extending the previously developed 2D framework into 3D, this work overcomes the inherent limitations of 2D simulations. 
Incorporating THB-splines enables efficient handling of multi-level local refinement.
Together with 3D dynamic domain expansion, the model allows for a high-resolution representation of neuron boundaries while keeping stable regions at low resolution for computational efficiency. 
The enhanced model delivers an accurate and biomimetic representation of 3D neurite morphologies, providing a tool for exploring neurodevelopmental processes and disorders.
By adjusting key parameters within the model, we can simulate morphological deterioration often observed in NDDs, thereby enhancing our understanding of the effects of various factors during this deterioration.
These simulations offer a computational platform to explore the potential impact of specific biophysical factors on neurodevelopmental process in the context of NDDs.
These novel contributions make the 3D phase field neuron growth model a robust framework for investigating NDDs. 
Furthermore, its ability to provide neuron growth predictions holds great potential for applications in therapeutic intervention planning, supporting the development of targeted treatment strategies. 
The model represents a key step forward in bridging computational approaches with experiments, gaining a deeper understanding of complex neurological challenges.

The 3D IGA-based phase field model demonstrates notable improvements, but computational challenges remain. 
High-fidelity simulations in 3D still demand substantial computational resources and efficient parallelization strategies. 
Future work will integrate machine learning techniques to enhance the phase field neuron growth framework, potentially accelerating simulations and improving scalability for large-scale 3D scenarios.
While the model effectively incorporates key biophysical processes, simplifying assumptions limits its ability to fully replicate the intricacies of neurodevelopmental processes. 
% \textbf{mention $D_c k_{p75}$ in \ref{eqn: NDDs synaptogenesis_equation}, but we only studied $c_{opti}$, but there  are other coeficients can be studies, which can be future work. rewrite next part to use our model to get grown neurite, and then use the neurite morphology to study transport}
% Expanding the model to include more complex mechanisms, such as intracellular transport dynamics \cite{li2019modeling, li_deep_2021, li2022modeling_1, li2022modeling, li2023isogeometric} and neuron-environment interactions, will improve its predictive accuracy and applicability to real-world biological systems.
Eq.~\ref{eqn: NDDs synaptogenesis_equation} also includes $D_c$ and $k_{p75}$ that affect the distribution of $c_{neur}$, our current study focuses on $c_{opti}$ to investigate retraction and deteriorations. 
Investigating the influence of other coefficients on neurite growth is a compelling future research direction.
Using our developed 3D phase field model, we can simulate the growth of a neurite under specific conditions. 
The resulting neurite morphology can then be analyzed to study transport phenomena~\cite{li2019modeling, li_deep_2021, li2022modeling_1, li2022modeling, li2023isogeometric} and neuron-environment interactions, which will improve its predictive accuracy and applicability to real-world biological systems.

Ongoing efforts aim to further refine the model by incorporating additional biological complexities and evaluating its performance under diverse conditions. 
We plan to simulate neurotrophin gradients, competitive tubulin consumption, and extracellular matrix interactions within a biophysics-based 3D culture environment with varying substrates. 
Validations will involve direct comparison with experimental data, ensuring the model is accurate and reliable in replicating observed growth patterns.

\section*{Code and data availability}
The code and datasets generated and analyzed in this paper are accessible in the "3D\_THBNG" GitHub repository. \href{https://github.com/CMU-CBML/3D_THBNG}{\nolinkurl{https://github.com/CMU-CBML/3D_THBNG}} (\href{https://doi.org/10.5281/zenodo.15331690}{doi:10.5281/zenodo.15331690}). 
Videos are also available in the github folder. Correspondence and requests for code and data should be addressed to K.Q. or Y.J.Z.

\section*{Declaration of competing interest}
The authors declare no known competing financial interests or personal relationships that could have appeared to influence the work reported in this paper.

\section*{Acknowledgement}
K. Qian and Y. J. Zhang were supported in part by the NSF grants CMMI-1953323 and CBET-2332084.
% V. A. Webster-Wood was supported in part by an NSF CAREER award ECCS-2044785. 
This work used RM-nodes on Bridges-2 Supercomputer at Pittsburgh Supercomputer Center \cite{ecss,xsede} through allocation ID eng170006p from the Advanced Cyberinfrastructure Coordination Ecosystem: Services \& Support (ACCESS) program, which is supported by National Science Foundation grants \#2138259, \#2138286, \#2138307, \#2137603, and \#2138296.

\bibliographystyle{elsarticle-num}
\bibliography{reference}
\end{document}